\documentclass{svproc}

\usepackage{url}
\usepackage{siunitx}
\usepackage{pgfplots}

\begin{document}
\mainmatter              

\title{Influence of Thermostats on the Dynamics of the Helix-Coil Transition}

\titlerunning{Dynamics of the Helix-Coil Transition}  
\author{Maximilian Conradi\inst{1,*} \and Henrik Christiansen\inst{2} \and Suman Majumder\inst{3} \and \\Fabio Müller\inst{1} \and Wolfhard Janke \inst{1}}

\authorrunning{Conradi et al.} 
\tocauthor{Maximilian Conradi, Wolfhard Janke, Henrik Christiansen, Suman Majumder, Fabio Müller}

\institute{Institut für Theoretische Physik, Universität Leipzig, IPF 231101, 04081 Leipzig, Germany\\
\and
NEC Laboratories Europe GmbH, Kurfürsten-Anlage 36, 69115 Heidelberg, Germany
\and
Amity Institute of Applied Sciences, Amity University Uttar
Pradesh, Noida 201313, India}
\maketitle              

\begin{abstract}
We present results from all-atom molecular dynamics simulations for the nonequilibrium dynamics of the collapse and helix-coil transition in polyalanine.
In particular, we compare the influence of three different thermostats, viz., the Langevin, Andersen, and Nosé-Hoover thermostats.
For that purpose, we investigate the nonequilibrium pathways of the transition from the high-temperature random-coil state to the low-temperature helical state.
Additionally, we analyze the time evolution of the potential energy and temperature.
Our results show only small differences in the observed phenomenology, albeit quantitatively the dynamics appear to be different for the three thermostats.
{\let\thefootnote\relax\footnote{* Email: \email{conradi@itp.uni-leipzig.de}}} 

\end{abstract}

\section*{Introduction}

Proteins are responsible for and influence most biological processes, making understanding the protein biosynthesis crucial \cite{dob,dil:cal}.
While some aspects of the biosynthesis are well understood, the understanding of some important details is still lacking.
Specifically, the mechanisms related to the formation of secondary structure elements are not completely understood.
Among the many ways to investigate this problem, computer simulations are one of the most promising ones \cite{dil:cal,dil:ozk}.
In particular, molecular dynamics (MD) simulations \cite{fre:smi} allow for a detailed study of folding trajectories.
However, simulating a full protein can be computationally very expensive due to the large number of atoms -- especially in presence of an explicit solvent.
As a way out, previous studies investigated the kinetics of the collapse transition of simple coarse-grained polymers as a prerequisite to folding of proteins \cite{byr:kie,chr:maj,maj:zie}. 

In such collapse studies one monitors the nonequilibrium pathways from an initially 
disordered random-coil state at high temperature to the collapsed state after
a sudden quench to low temperature. For coarse-grained polymers the dynamics
of the collapse transition is slower than for proteins. A specific example 
is the biopolymer polyglycine for which a recent study observed a transition 
that is significantly faster than that in the coarse-grained formulation \cite{maj:han:jan}. 

Beside these differences based on the type of polymer, differences have
also been found when comparing different simulation methods. Generally the
dynamics is faster in MD compared to Monte Carlo simulations \cite{maj}. In view of
the above discussion, it is crucial to investigate various simulation methods for
a better comparability when studying the collapse kinetics. Since
in such studies the temperature is an important control parameter, we here focus 
on the influence of the thermostat in an MD simulation on the concerned
nonequilibrium process. 

As a model we use polyalanine, which has been shown to form helical structures in solvent at low temperature \cite{Ara:Han}.
It has a relatively simple structure and specifically its helix-coil transition makes it intriguing.

\section*{Methods}
We consider three different thermostats: Langevin (LT), Andersen (AT), and Nosé-Hoover chain (NHT).
Thermostats can be generally categorized into stochastic and deterministic thermostats, with LT and AT belonging to the family
of stochastic and NHT belonging to the family of deterministic thermostats.

LT models a system in contact with a heat bath at temperature $T$ and propagates the system by means of the Langevin equation
\begin{equation}
	m_i\ddot{\vec{r}}_i = \vec{f}_i - \gamma m_i\dot{\vec{r}}_i + \sqrt{2m_i\gamma k_BT} \vec{R}_i.
\end{equation}
Here $\dot{\vec{r}}_i=d\vec{r}_i/dt$ and $m_i$ denote, respectively, the velocity and mass of a particle, $\vec{f}_i$ the conservative
force acting on it, $\gamma$ the friction coefficient taking care of the drag force due to the solvent, and the last term accounts for
thermal fluctuations with $\vec{R}_i$ drawn from a normal distribution with mean zero and unit variance, being $\delta$-correlated over different
particles and time \cite{hue}. 

Similarly, also in AT the system is coupled to a heat bath. Here, after each MD step, a number of particles is randomly chosen with probability 
$P(t)=\nu \Delta t$, where $\Delta t$ is the time step and $\nu$ is the collision frequency. The selected particles collide with the heat-bath 
particles and are then assigned a new velocity  $\dot{\vec{r}}^*_i=\sqrt{k_B T/m}\vec{R}_i$ from the Maxwell-Boltzmann distribution at temperature 
$T$, with $\vec{R}_i$ as Gaussian random numbers with mean zero and unit variance \cite{and}.
This results in the equation of motion
\begin{equation}
	\ddot{\vec{r}}_i=m_i^{-1}\vec{f}_i + \sum_{n=1}^{\infty}\delta\left(t -\sum_{k=1}^n\tau_{i,k}\right)[\dot{\vec{r}}_{i,n}^*(t)-\dot{\vec{r}}_i(t)]
\end{equation}
where $\tau_{i,k}$ is the series of intervals in which particle $i$ does not get assigned a new velocity and $\dot{\vec{r}}_{i,n}^*$ is the randomly assigned new velocity \cite{hue}.

NHT is based on an extended system \cite{nos} with additional degrees of freedom $\eta_j$ and corresponding masses $Q_j$.
Here, the first thermostat $\eta_1$ controls the fluctuations of the system.
The other thermostats $\eta_j$ are then used to control the fluctuations of the preceding thermostat $\eta_{j-1}$, creating a chain of thermostats \cite{nos,hoo,mar:kle:tuc}.
The dynamics of the system at temperature $T$ is then given by the following equations of motion:

\begin{eqnarray}
\dot{\vec{r}}_i = \frac{\vec{p}_i}{m_i}, 
&\,\,\,&
\dot{\vec{p}}_i = \vec{f}_i\, - \vec{p}_i \frac{p_{\eta_1}}{Q_1}, \qquad \nonumber \\ 
\dot{p}_{\eta_1} = \left[\sum_{i=1}^N \frac{\vec{p}_i^{\,2}}{m_i}-3Nk_BT\right] - p_{\eta_1} \frac{p_{\eta_2}}{Q_2},
&\,\,\,&
\dot{p}_{\eta_j} = \left[\frac{p_{\eta_{j-1}}}{Q_{j-1}} -k_BT\right] - p_{\eta_j} \frac{p_{\eta_{j+1}}}{Q_{\eta_{j+1}}},
\qquad
\label{eq:NHT}
\end{eqnarray}

together with the equations $\dot{\eta_j} = p_{\eta_j}/Q_j$ for the heat-bath ``coordinates`` $\eta_j$, $j=1,\dots,M$, which are decoupled from 
the dynamics.

In contrast to NHT neither LT nor AT are deterministic due do the inclusion of random elements in the interaction with the medium. 
LT and NHT are believed to realistically reproduce the dynamics of the system by incorporating solvent effects.
This is to some extent because LT includes a drag force to replicate the interactions with the solvent and NHT conserves the linear momentum 
and has been shown to preserve hydrodynamic effects \cite{hue,maj2013}.

In order to study these influences we use polyalanine as a model. Each molecule consists of $N= 100$ residues and is capped with a hydrogenated N-terminus ($-NH_2$) and C-terminus ($-COOH$).
Using the OpenMM package \cite{openMM} the molecule is first equilibrated at a high temperature of $T=\SI{2000}{\kelvin}$ in a generalized Born implicit solvent \cite{ngu:roe}.
The sampled conformations are then quenched to a temperature of $T=\SI{300}{\kelvin}$, where we use the Amber14ffSB force field and the generalized Born implicit solvent for simulation.
We take $\Delta t=\SI{1}{\femto\second}$, $\gamma = \SI{e-3}{\per\femto\second}$, and $\nu = \SI{e-3}{\per\femto\second}$, respectively, for the thermostats. Averaging is performed over at least 100 independent starting realizations.
All simulations were performed on a GPU cluster.

\section*{Results}
\begin{figure}[t!]
\begin{center}
\begin{minipage}{.06\linewidth}
LT
\end{minipage}
\begin{minipage}{.225\linewidth}
\centering
\includegraphics[width=\linewidth]{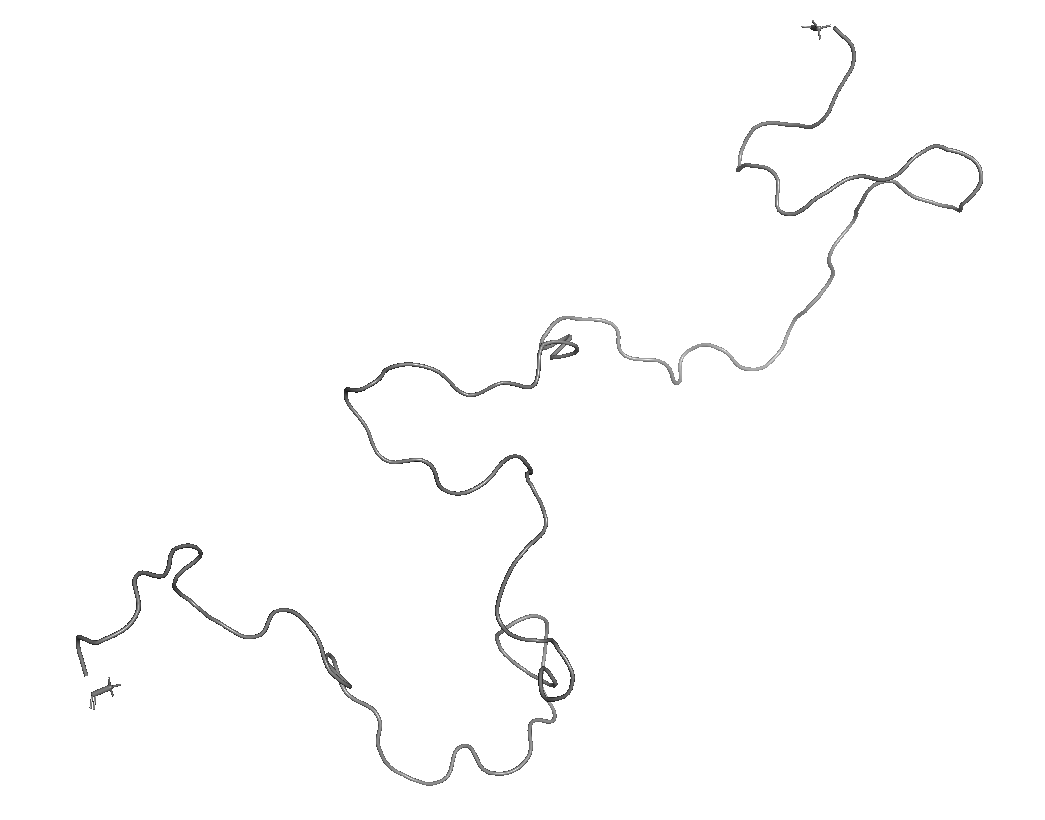}
\end{minipage}
\hfill
\begin{minipage}{.225\linewidth}
\centering
\includegraphics[width=\linewidth]{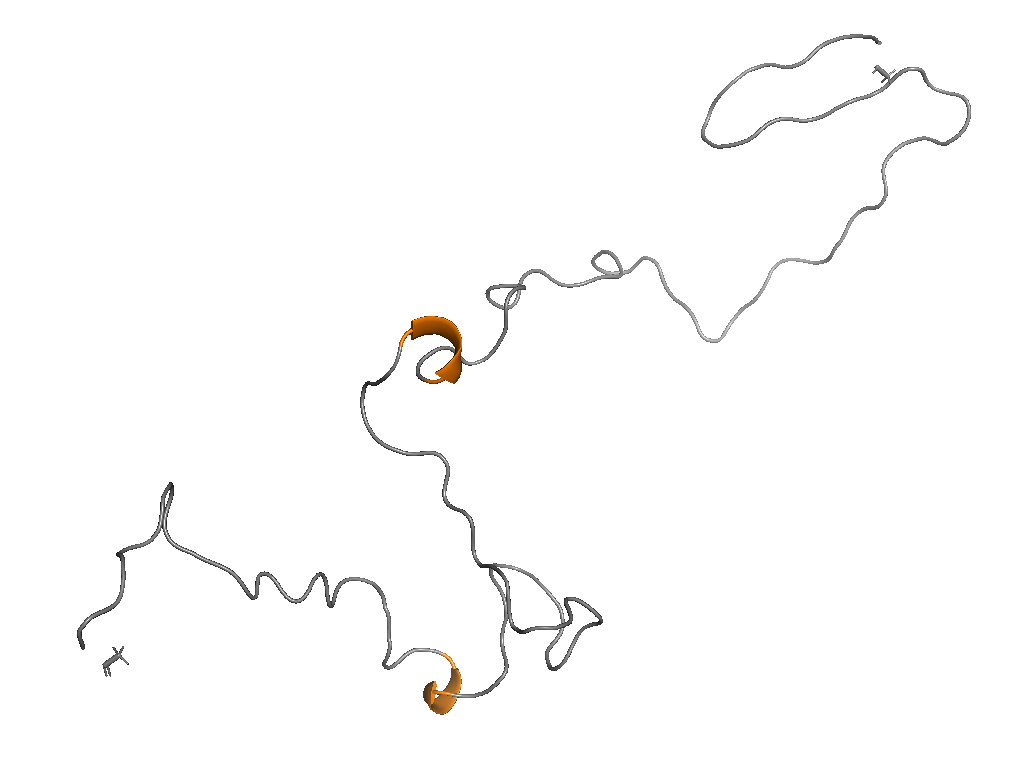}
\end{minipage}
\hfill
\begin{minipage}{.225\linewidth}
\centering
\includegraphics[width=\linewidth]{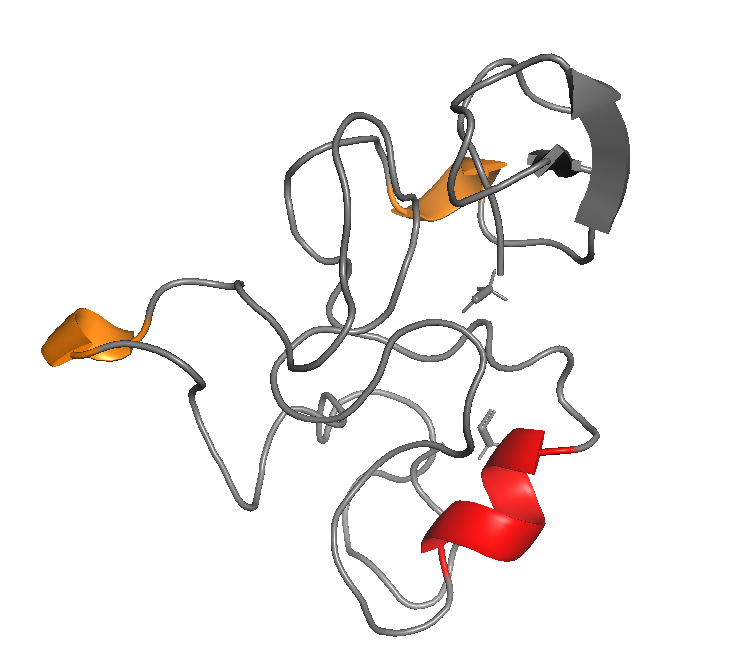}
\end{minipage}
\hfill
\begin{minipage}{.225\linewidth}
\centering
\includegraphics[width=\linewidth]{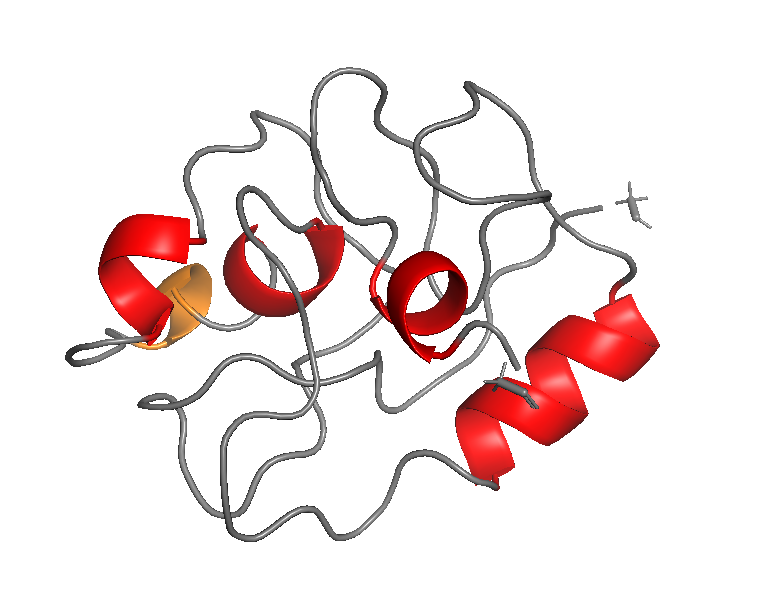}
\end{minipage}
\vspace{.1cm}
\begin{minipage}{.06\linewidth}
AT
\end{minipage}
\begin{minipage}{.225\linewidth}
\centering
\includegraphics[width=\linewidth]{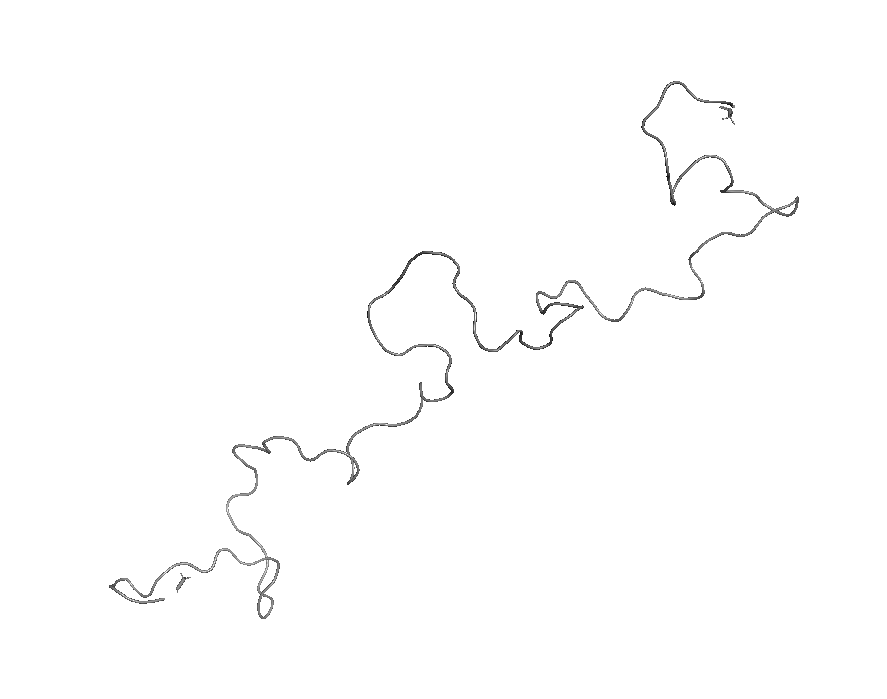}
\end{minipage}
\hfill
\begin{minipage}{.225\linewidth}
\centering
\includegraphics[width=\linewidth]{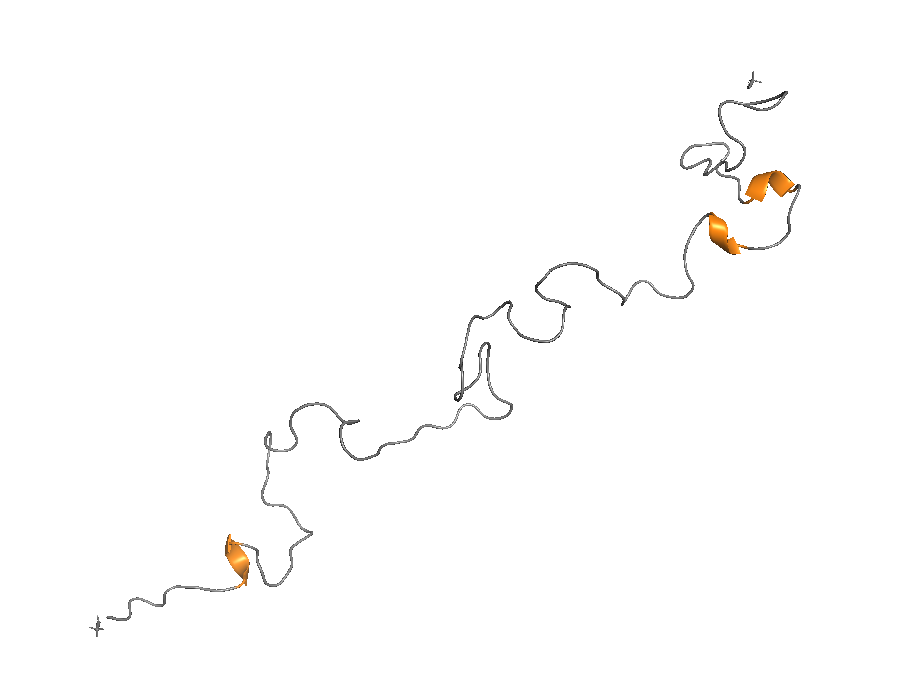}
\end{minipage}
\hfill
\begin{minipage}{.225\linewidth}
\centering
\includegraphics[width=\linewidth]{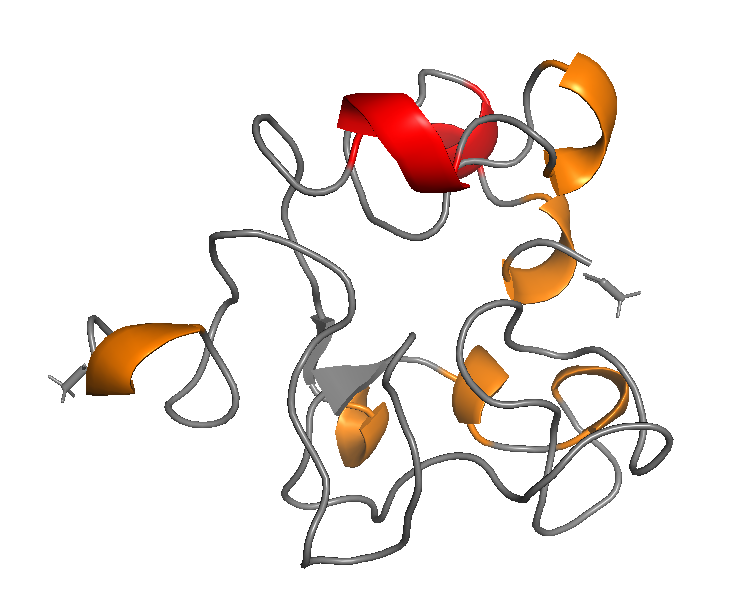}
\end{minipage}
\hfill
\begin{minipage}{.225\linewidth}
\centering
\includegraphics[width=\linewidth]{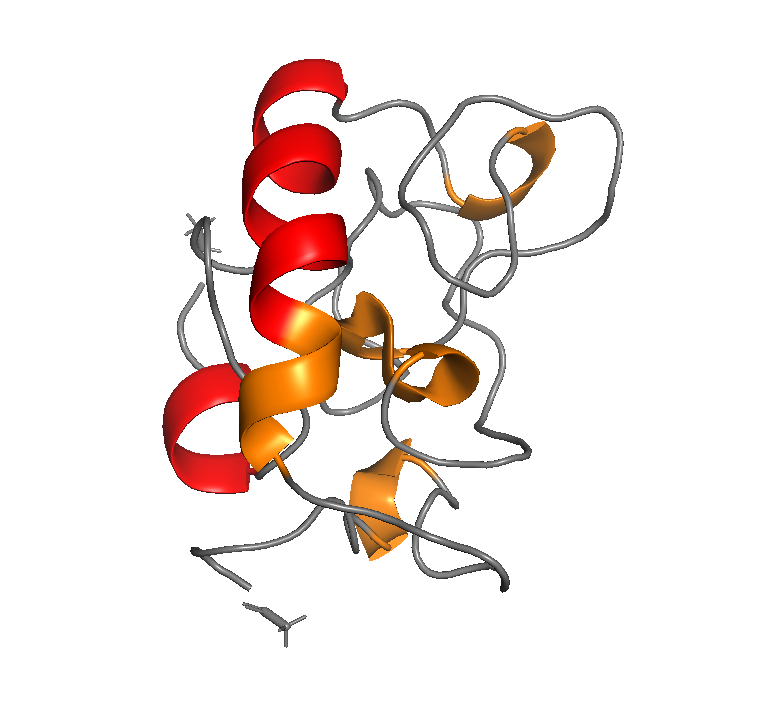}
\end{minipage}
\vspace{0.1cm}
\begin{minipage}{.06\linewidth}
NHT
\end{minipage}
\begin{minipage}{.225\linewidth}
\centering
\includegraphics[width=\linewidth]{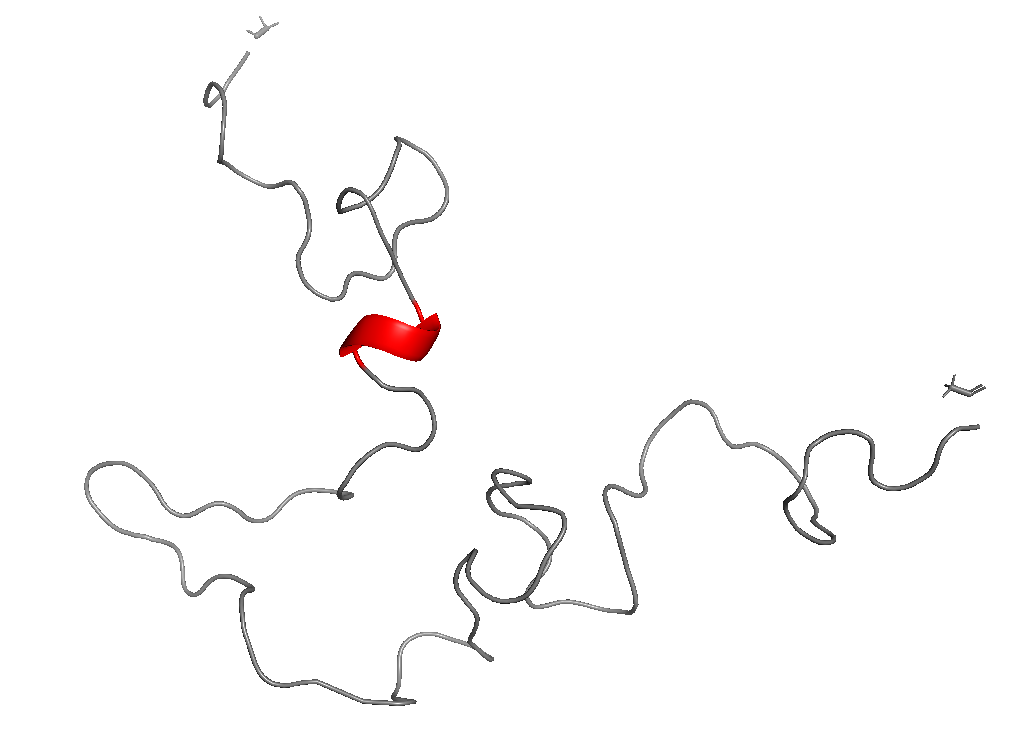}
\end{minipage}
\hfill
\begin{minipage}{.225\linewidth}
\centering
\includegraphics[width=\linewidth]{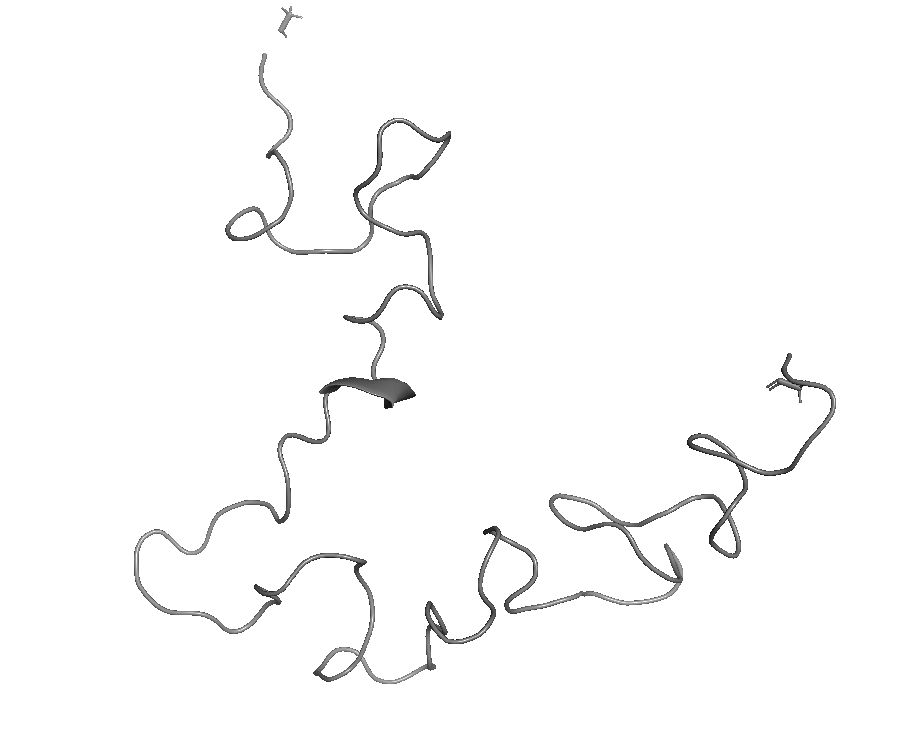}
\end{minipage}
\hfill
\begin{minipage}{.225\linewidth}
\centering
\includegraphics[width=\linewidth]{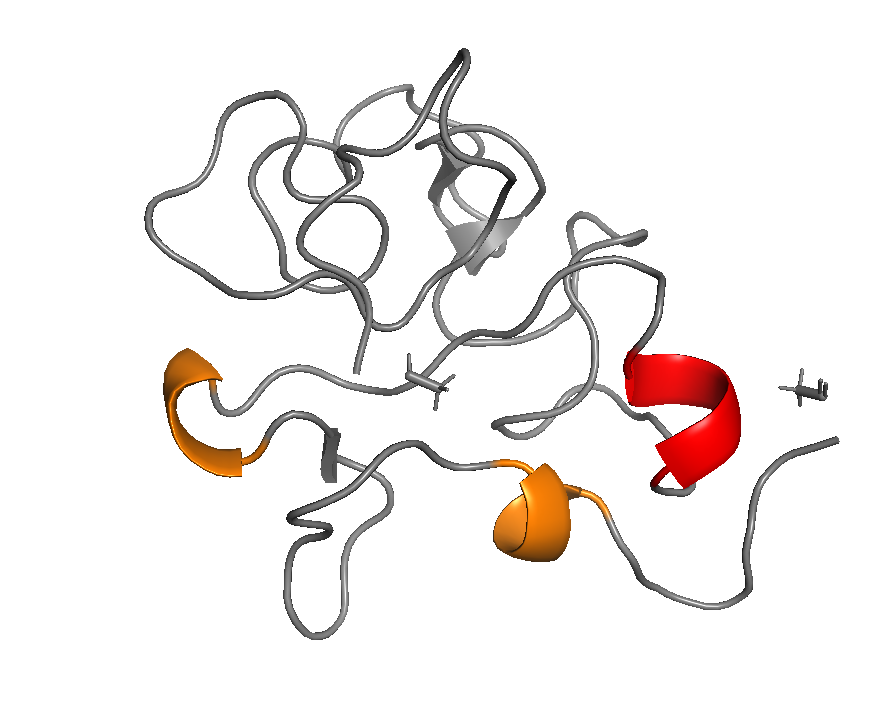}
\end{minipage}
\hfill
\begin{minipage}{.225\linewidth}
\centering
\includegraphics[width=\linewidth]{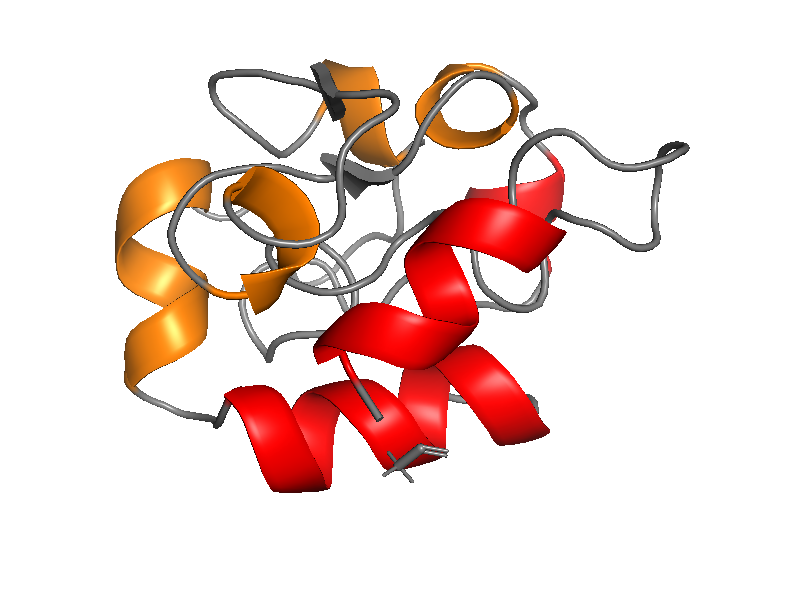}
\end{minipage}
\vspace{0.1cm}
\begin{minipage}{.06\linewidth}
\end{minipage}
\begin{minipage}{.225\linewidth}
\centering
$t=\SI{0}{\nano\second}$
\end{minipage}
\hfill
\begin{minipage}{.225\linewidth}
\centering
$t=\SI{0.01}{\nano\second}$
\end{minipage}
\hfill
\begin{minipage}{.225\linewidth}
\centering
$t=\SI{1}{\nano\second}$
\end{minipage}
\hfill
\begin{minipage}{.225\linewidth}
\centering
$t=\SI{100}{\nano\second}$
\end{minipage}
\end{center}
\label{traj}
\caption{Snapshots during the transition for the three thermostats. $\alpha$-helical residues are colored red, $3_{10}$-helical residues are colored orange. The arrow represents a strand in a $\beta$-sheet. Secondary structure elements are determined using DSSP \cite{kab}.}
\end{figure}

Figure\ \ref{traj} shows exemplary snapshots from one trajectory for each of the three thermostats.
The molecules are in an extended state before the quench at $t = \SI{0}{\nano\second}$.
In the second image of each row we see that there is not much progression in both AT and NHT until $t = \SI{0.01}{\nano\second}$.
Meanwhile, the molecule simulated with LT has already slightly shortened and formed small clusters in some areas.
After $\SI{1}{\nano\second}$, all molecules have collapsed to a globule and have already formed some helical structures.
Finally, after $\SI{100}{\nano\second}$  in all cases the molecule has a compact shape with longer helical segments.
Based on these trajectories, the transition appears to be split into a phase that is dominated by the collapse from the random-coil state to a
globular state and a second phase, which is dominated by the formation of helical structures. As a result the molecule becomes more extended.
This is in agreement with earlier studies \cite{Polyala} which also found this separation into these two separate processes.

\begin{figure}[t!]
	\begin{minipage}{.49\linewidth}
	\centering
	\includegraphics[width=\linewidth]{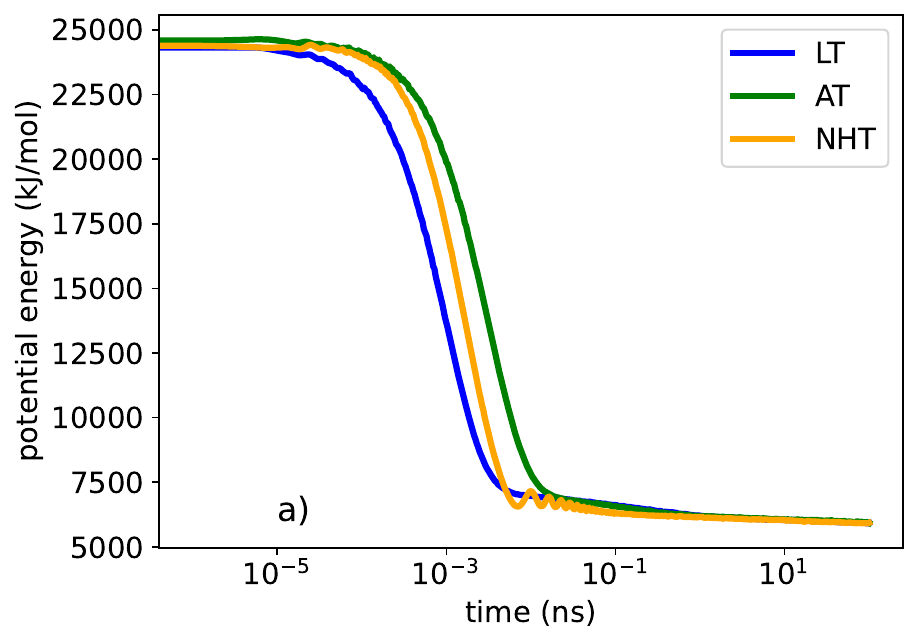}
	\end{minipage}
	\hfill
	\begin{minipage}{.49\linewidth}
	
	\includegraphics[width=\linewidth]{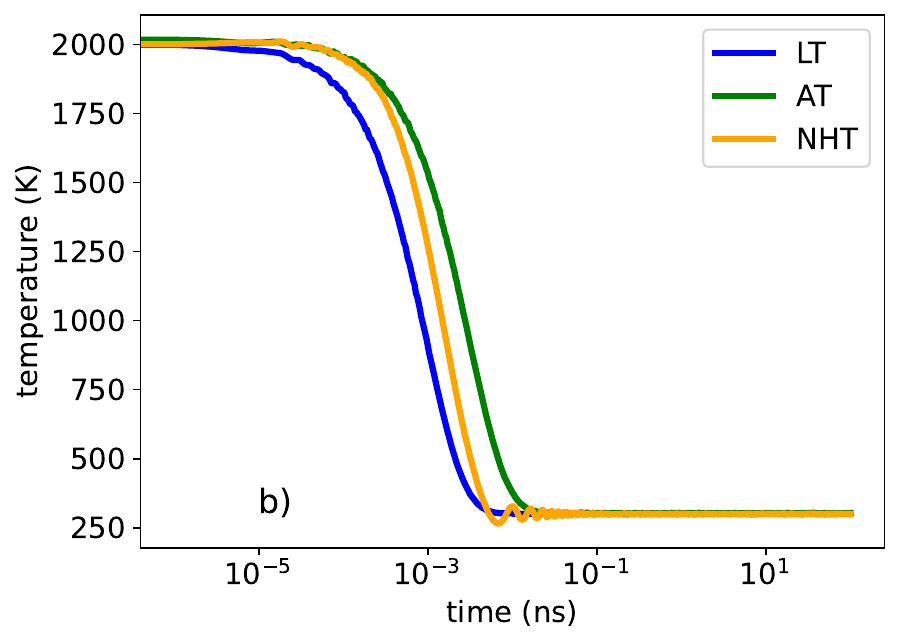}
	\end{minipage}
	\caption{Time evolution of a) the potential energy and b) the (effective) temperature during the helix-coil-transition. The curves show averages over 100 independent trajectories. Error bars are too small to be visible in the plots.}
	\label{temp}
\end{figure}

The faster collapse with LT is also reflected in the time evolution of the potential energy shown in Fig.\ \ref{temp}a).
Here, we observe that already after only around \SI{e-4}{\nano\second} the potential energy starts decreasing.
For AT and NHT, this decrease sets in at approximately \SI{e-3}{\nano\second}.
NHT shows a very sharp decrease, while AT and LT have a very similar but less sharp decrease.
In all cases the decline is rapid and ends after about \SI{0.01}{\nano\second}.
At this point, one can see that for NHT in the potential energy small oscillations occur, which are probably an artifact of the thermostat \cite{hue}.
Following this, the potential energy decreases very slowly in the second stage.
The rapid decrease corresponds to the collapse of the polymer and the slow regime to the formation of helices.

A comparison with the effective temperature, determined from the kinetic energy, shown in Fig.\ \ref{temp}b) yields almost the same picture.
With LT, the system starts adjusting its temperature at the earliest time and also reaches the final temperature first.
While the decrease in temperature sets in later for the other thermostats, the adjustment is faster with NHT and settles at around the same time as LT.
Again, one observes the fluctuations in NHT, when reaching the new equilibrium. During the helix formation the temperature seems to be constant. 

The visible similarity between the trajectories of LT and AT could be explained by the stochastic nature of both thermostats.
The later start of the collapse with AT can be explained by the fact that AT only adjusts single particles while LT acts on all particles.
The low collision frequency of $\nu=\SI{e-3}{\per\femto\second}$ then causes the collapse to start relatively late.

\section*{Conclusion}
In conclusion, we have shown that the thermostat for an MD simulation should be chosen with great care, as it influences the nonequilibrium dynamics of the system.
Our comparative study of three thermostats shows that following a temperature quench the potential energy decreases at different rates.
NHT shows a faster adjustment of the temperature than AT and LT.
In general the transition appears to be split into a first stage dominated by the collapse of the random-coil state and a second stage showing mainly the formation of helical structures. 
 
While the qualitative differences between the thermostats are small in this analysis they still should not be ignored. Already the faster onset of the collapse 
with one of the thermostats could influence results in more quantitative analyses.
However, besides the nature of the thermostat, parameters controlling the thermostat may also influence the dynamics of the system.  
This has indeed been demonstrated for studying critical dynamics of a vapor-liquid system in equilibrium \cite{roy:das}. This could be crucial for the helix-coil transition as well.

\section*{Acknowledgments}
This project was funded by the Deutsche Forschungsgemeinschaft (DFG, German Research Foundation) -- 469\,830\,597 under project ID JA 483/35-1. S.M. thanks the Science and Engineering Research Board (SERB), Govt.\ of India for a Ramanujan Fellowship (file no.\ RJF/2021/000044). All simulations were performed on the GPU cluster of the Universitätsrechenzentrum (URZ) at Universität Leipzig.

\end{document}